\documentclass[9pt,twocolumn,twoside]{osajnl}
\journal{ol} 
\usepackage{tikz}
\usetikzlibrary{arrows.meta,decorations.pathreplacing,shadows}
\graphicspath{{./figures/}}

\setboolean{shortarticle}{true} 

\title{Electro-optically modulated lasers: important pulse energy and contrast optimization derived from theoretical spectrum}
\author[1,*]{Denis Marion}
\author[1]{Jérôme Lhermite}

\affil[1]{Universit\'e de Bordeaux, CNRS, CEA, CELIA (Centre Lasers Intenses et Applications), UMR 5107, 33405 Talence, France}
\affil[*]{Corresponding author: denis.marion@u-bordeaux.fr}

\dates{31 March 2020}

\begin{abstract}
In this article, we derive analytically the complex optical spectrum of a pulsed laser source obtained when a frequency comb generated by phase modulation is input into a synchronized intensity modulator. 
As the exact knowledge of the spectrum allows to reshape it arbitrarily with an optical spectrum processor, we show how this operation can lead to a 75\% increase in pulse energy and an enhancement of the  extinction ratio by at least three orders of magnitude in some cases. This method also enables large-factor rate-multiplications of these versatile coherent sources using the Talbot effect with negligible degradation of the signal.
\end{abstract}

\doi{\url{https://dx.doi.org/10.1364/OL.391192}}
\setboolean{displaycopyright}{true}

\begin{document}
\maketitle
\thispagestyle{fancy}

Electro-optically modulated laser (EOML) setups are an emerging class of pulsed laser sources delivering short light bursts at an ultra-high repetition rate (RR, typically tens of GHz), usually tunable up to one or more octaves. Contrarily to cavity-based mode-locked laser oscillators, these sources rely on one or more phase modulators (PM) driven by a radio-frequency (RF) oscillator to generate a wide optical spectrum. They are, in that sense, very similar to electro-optic frequency combs. However, although it was not the case in the earlier designs  \cite{kobayashi88}, a synchronized intensity modulator (IM) is nowadays generally added, which is not the case for frequency combs used in the spectral domain. 

The spectral phase of an EOML is directly linked to the synchronization of all the phase and intensity modulators. Neglecting the phase noise of the RF oscillator as well as synchronization errors, it is definite, constant and even \cite{metcalf13}. As a direct consequence, the pulses emitted by these sources can readily be compressed very close to their Fourier time-limited (FTL) duration. Recent EOMLs deliver very short pulses ($\leqslant 350$~fs) without the need for a nonlinear mode-locking process.  

Historically, the first developments were carried out around 1550~nm, where wideband RF modulators based on lithium niobate waveguides were first available  \cite{murata2000,ishizawa2011,ishizawa2013,wu2010a,wang2011a}, but more recently new setups centered around 1~\textmu m were demonstrated  \cite{Aubourg:15,metcalf2019}, giving access to the high average power of ytterbium-doped fiber amplifiers (YDFA) and also allowing the nonlinear post-compression of the pulses down to \cite{metcalf2019} 80~fs. 

EOMLs have some drawbacks: they are relatively expensive as their total spectral width is dictated by the number of PMs, and their extinction ratio over the RR period, defined here as the peak-to-valley instantaneous power ratio, is orders of magnitude higher than that of typical mode-locked lasers. However, EOMLs gather many unique features. They are reliable all-fiber sources without any need for nonlinear component and offer a very large span of repetition rate in an unusual range (larger than 10~GHz typ.). Being based on low spectral width laser emitters, they can also easily be tuned to the desired central wavelength by changing the emitter and/or fine-tuned by adjusting its temperature. Their spectral properties can be readily simulated, as no population inversion nor nonlinear propagation needs be described  \cite{Aubourg:15}. A more detailed review of technologies related to ultrahigh repetition rate sources and electro-optic combs can be found elsewhere \cite{Torres2014} and is beyond the scope of this work. 

The IM enhances the contrast over the period by at least one order of magnitude, but it also modifies the spectrum of the frequency comb in a non negligible way. Note that, to the best of our knowledge, such an analysis of an EOML mixing phase and intensity modulations has never been presented before. 

The first part of this work is dedicated to the analytical calculation of the complex spectrum. We then use the derived expressions to analyze the spectral phase and show that a pulse shaper is generally essential to avoid detrimental artefacts in the time domain. Lastly, we show how this derivation can be used, in conjunction with such a device, to optimize greatly the time profile and make it suitable for rate-multiplication by temporal Talbot effect. Throughout this paper, we use the term "pulse shaper" to designate an optical spectrum processor \citep{Roelens:08} based on a 2D spectral light modulator (SLM) allowing to shape both the amplitude and the phase of the optical spectrum. Nowadays, such devices with a spectral resolution of typically 20 GHz or less are commercially available.

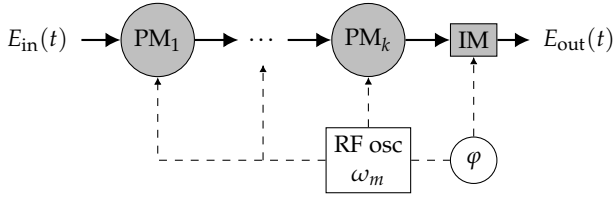
\begin{figure}[ht]
\centering
\begin{tikzpicture}[scale=0.8,node distance=1.5cm, every node/.style={fill=white}, align=center,>={Latex[width=2mm,length=2mm]}]
\tikzstyle{PM}=[draw,circle,fill=black!25];
\tikzstyle{IM}=[draw,rectangle,fill=black!25,minimum width=1];
\tikzstyle{field}=[circle,fill=white];

\tikzstyle{transf}=[style={thick, ->}];
\tikzstyle{RF}=[style={dashed, ->, >={Latex[width=1mm,length=1mm]}}];

\node[field] (E0) at (-.25,0) {$E_\text{in}(t)$};
\node[PM] (PM1) at (1.75,0) {PM${}_1$}; 
\node[field] (E1) at (3.5,0) {\dots};
\node[PM] (PMk) at (5.25,0) {PM${}_k$}; 
\node[IM] (IM) at (7,0) {IM}; 
\node[field] (Eout) at (8.75,0) {$E_\text{out}(t)$}; 

\node[IM, fill=white] (osc) at (5.25,-2) {RF osc \\$\omega_m$};
\node[PM, fill=white] (phi) at (7,-2) {$\varphi$};

\draw[transf] (E0) -- (PM1) ;
\draw[transf] (PM1) -- (E1) ;
\draw[transf] (E1) -- (PMk) ;
\draw[transf] (PMk) -- (IM) ;
\draw[transf] (IM) -- (Eout);

\draw[RF] (osc) -- (1.75,-2) -- (PM1); 
\draw[RF] (3.5,-2) -- (E1);
\draw[RF] (osc) -- (PMk);
\draw[RF] (osc) -- (phi) -- (IM);

\end{tikzpicture}

\caption{\label{fig:EOML setup} Generalized description of an EOML setup. PM${}_j$:~synchronized phase modulators, RF osc: radio-frequency oscillator delivering a perfect sine law of pulsation $\omega_m$. The IM phase relative to the PMs is $\varphi$.}
\end{figure}
 
We restrict our approach to the EOML systems equivalent to that depicted in Fig. \ref{fig:EOML setup}. All PMs are supposed synchronized. We also make the hypothesis that the optical electric field can be described by scalar theory.

The input field $E_\text{in}(t)$ is usually generated by a quasi-monochromatic laser diode. Here, it is supposed of the form $E_\text{in}(t) = E_0 e^{\imath \omega_0t}$ where $\omega_0 = 2\pi c/\lambda_0$ is the optical pulsation of the diode and $c$ is the speed of light in vacuum. PM${}_j$ adds an instantaneous phase modulation by the quantity 
\begin{equation}
\label{Eq: phi_j}
\phi_j(t) = \pi\frac{V_j}{V_{\pi}}\sin\omega_mt
\end{equation}
where $V_j$ is the amplitude of the voltage on PM${}_j$ and $V_{\pi}$ their half-wavelength voltage, supposed common here for all the modulators (PMs and IM). $\omega_m=2\pi/T_m$ is the modulation pulsation of all modulators. Since the PMs of the setup have a null phase relative to the RF oscillator, the scalar electric field at the output of PM${}_k$ reads
 \begin{equation}
 \label{Eq:E_PM}
 E_\text{PM}(t) = E_0 e^{\imath \omega_0t} e^{\imath (\phi_1(t) + \dots +  \phi_k(t))} = E_0 e^{\imath \omega_0t} e^{\imath \beta \sin{\omega_m t}}
 \end{equation}
where $\beta$ is the total phase modulation depth of the EOML:
 \begin{equation}
 \label{Eq:beta_def}
 \beta = \pi \sum_{j=1}^{k}\frac{V_j}{V_{\pi}}
 \end{equation}

The IM is usually a miniaturized Mach-Zehnder modulator embedded on a lithium niobate waveguide. We may assume without loss of generality that its amplitude transmission is in the form
\begin{equation}
\label{Eq:transmission IM}
\tau(t) = \sin\left[\Phi_0 + \Phi_1 \sin (\omega_mt + \varphi)\right]
\end{equation}
where $\Phi_0$ is an offset controlled via the bias voltage of the IM and $\Phi_1$ is the amplitude of modulation given by $\Phi_1 = V_\text{IM}/V_{\pi}$.

We recall the general Jacobi-Anger relation which gives the Fourier decomposition of a purely phase-modulated signal:
\begin{equation}
\label{Eq:Jacobi-Anger}
e^{\imath m\sin(\omega t+ \psi)} = \sum_{n=-\infty}^{+\infty} 
J_n(m) e^{\imath n \psi} e^{\imath n \omega t}
\end{equation}
where $J_n$ is the $n$-th Bessel function of the first kind. For the sake of conciseness, in the following we omit the infinite summation bounds. Using this relation and after some algebraic manipulations, the Fourier transform of $\tau$ reads
\begin{equation}
\label{Eq:tau^}
\hat{\tau}(\omega) =  \sum_{n}\alpha_n(\Phi_0,\varphi) J_n(\Phi_1) \delta(\omega - n\omega_m)
\end{equation}
where $\delta$ is the Dirac distribution and $\alpha_n(\Phi_0,\varphi)$ is defined as
\begin{align}
\label{Eq: alpha_n}
\alpha_n(\Phi_0,\varphi) = \begin{cases}
							\sin\Phi_0 e^{\imath n\varphi} \ &\text{if $n$ is even} \\
							-\imath\cos\Phi_0 e^{\imath n\varphi} \ &\text{if $n$ is odd}
							\end{cases}
\end{align}

The scalar field at the output of the EOML writes $E_\text{out}(t) = \tau(t)E_\text{PM}(t)$ or equivalently $\hat{E}_\text{out} = \hat{\tau} * \hat{E}_\text{PM}$, where $*$ denotes the convolution product. Using Eq. (\ref{Eq:Jacobi-Anger}) we obtain
\begin{equation}
\label{Eq:E^PM}
\hat{E}_\text{PM}(\omega) = E_0 \sum_n J_n(\beta) \delta_n(\omega)
\end{equation}
where $\delta_n(\omega) = \delta(\omega - \omega_0 - n\omega_m)$. From Eqs. (\ref{Eq:tau^}) and (\ref{Eq:E^PM}), the complex spectrum of an EOML can finally be written as
\begin{equation}
\label{Eq:E^out}
\hat{E}_\text{out}(\omega) = E_0\sum_{n} \gamma_n \delta_n(\omega)
\end{equation}
where 
\begin{equation}
\label{Eq:gamma_n}
\gamma_n = \sum_k\alpha_k(\Phi_0,\varphi) J_k(\Phi_1) J_{n-k}(\beta)
\end{equation}

Although the spectrum of EOMLs may be approximated by Fourier simulations, Eqs.~(\ref{Eq: alpha_n}) and (\ref{Eq:gamma_n}) give a clear understanding of its complex features. They also explicit the role of the IM and directly link its temporal parameters (bias, amplitude, relative phase) with the shape and phase of the spectrum. In the following, we name $E_\text{FL}$ the FTL function given by
\begin{equation}
\label{E_FL}
E_\text{FL} = E_0 \sum_n |\gamma_n| e^{\imath n \omega_mt}
\end{equation} 
$I_\text{FL}(t) = |E_\text{FL}(t)|^2$, and $\psi(\omega)$ the spectral phase $\arg({\hat{E}_\text{out}(\omega)})$.

We illustrate the rest of this work with the curves obtained for an ideal EOML ($\varphi = 0\ [\pi]$) modulated at $\beta = 20\pi$, a value easily reachable with current PMs and RF amplifiers \citep{Aubourg:15,metcalf2019}. Taking different values for $\beta$ alters the spectrum and temporal profile of EOMLS, but doesn't change the validity of the following matters.

\begin{figure}[ht]
\centering
\includegraphics[width=\columnwidth]{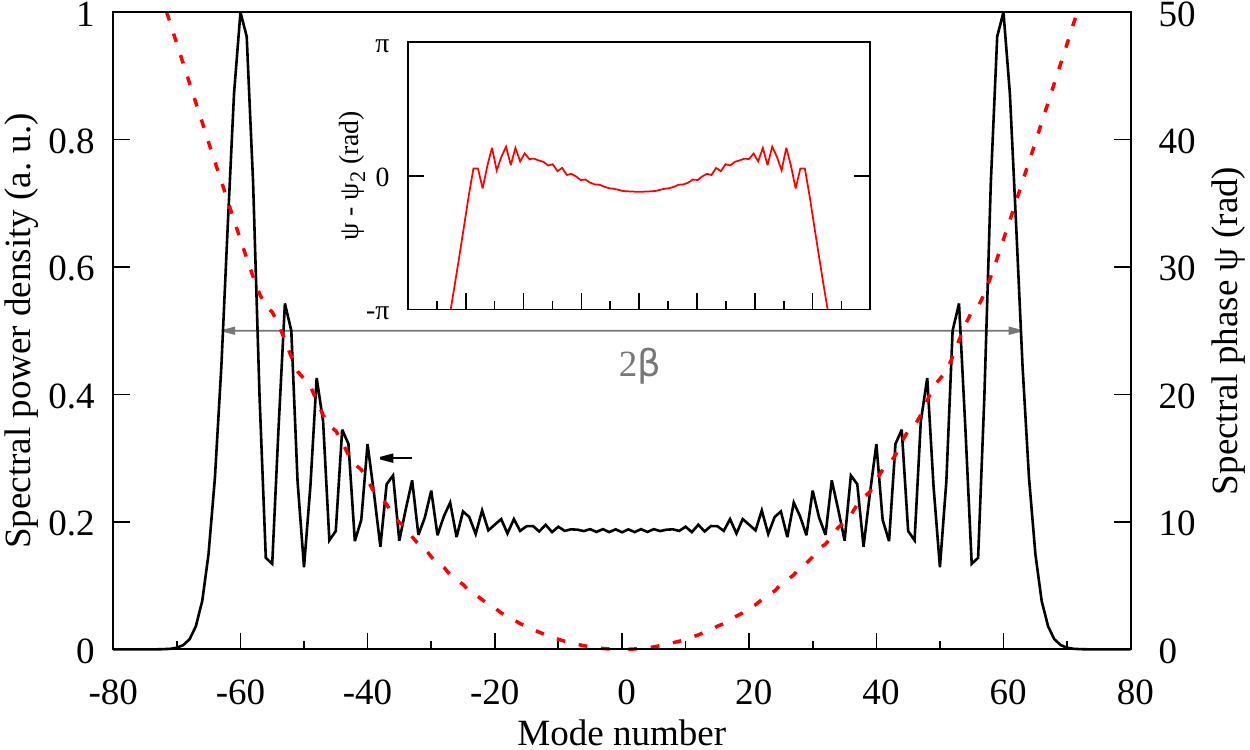}
\caption{\label{fig:SPD_phase_ideal}SPD (solid, black) and $\psi$ (dashed, red) of an ideal EOML ($\Phi_0 = \Phi_1 = \pi/4$, $\varphi = 0$). Inset: $\psi-\psi_2$.}
\end{figure}

As is clearly visible on Fig. \ref{fig:SPD_phase_ideal}, the FWMH of the spectrum is very well approximated by the value $2\beta$. Also, the unwrapped spectral phase of the ideal EOML is purely even. This is verified taking $\varphi = 0$ in Eq. (\ref{Eq: alpha_n}). Visually, the phase also appears to be mainly of $2^\text{nd}$ order. This is consistent with the successful compression setups based on a chirped volume Bragg grating (CVBG), a pulse shaper or a reflective grating in the literature  \citep{metcalf13,metcalf2019,marion20181,lhermite2019}. In order to emulate a $2^\text{nd}$-order compression, we carried out a least squares optimization with a polynomial of 2${}^\text{nd}$ degree. We name $E_2(t)$ the field obtained after subtracting the fitted phase $\psi_2$ from $\hat{E}_\text{out}$, and $I_2 = |E_2|^2$.

The remainder $\psi - \psi_2$ is presented in the inset of Fig. \ref{fig:SPD_phase_ideal}: the ripples clearly show that very high-order components are present in the spectral phase, which prevent a compression down to the FTL duration, with an effect both on the peak intensity and the pulse profile. Let us emphasize that static compressors such as reflective gratings or even tailored CVBGs are not an option in setups where $\beta$ and $\omega_m$ are adjustable, since $\psi$ and then $\psi_2$ change with these parameters.

\begin{figure}[ht]
\centering
\includegraphics[width=\columnwidth]{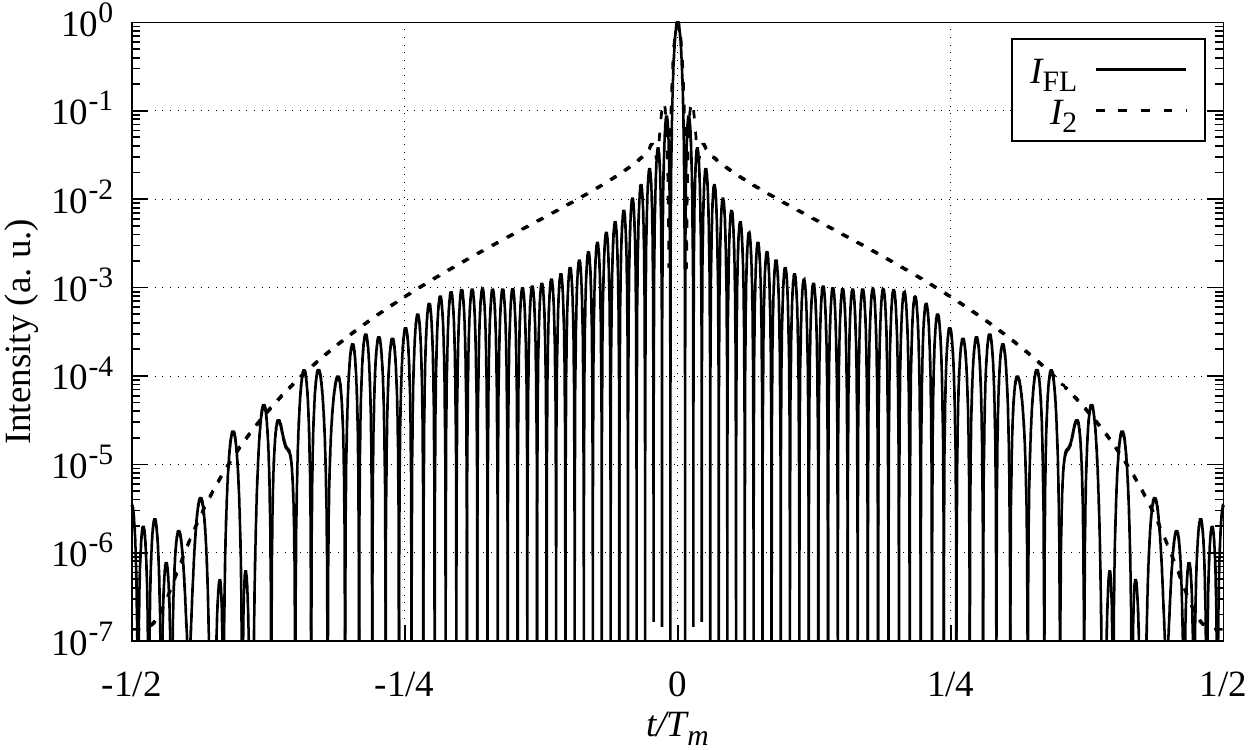}
\caption{\label{fig:eoml_contrast}Intensity of an EOML (scaled to 1 in magnitude) over one modulation period after a best-fit quadratic compression ($I_2$, dashed) or at Fourier limit ($I_\text{FL}$, solid).}
\end{figure}

Fig. \ref{fig:eoml_contrast} depicts the intensity over an entire period, for $E_\text{FL}$ as well as $E_2$. It illustrates the main drawback of EOMLs: their extinction ratio slowly reaches a plateau (here, about $10^{-3}$) one quadrant before and after the main pulse, and stays at $10^{-7}$ only on a small time range about one half-period after the pulse.

An important part of the residual phase can be corrected using a pulse shaper \cite{metcalf13,marion20181} provided $\omega_m/2\pi$ is superior to its frequency resolution. As the experimental determination of $\psi$ is complex, published setups rely on a step-by-step optimization of the programmed filter, hence leading to long procedures and only local optima. Alternatively, thanks to Eq.~(\ref{Eq:gamma_n}), the experimental SPD may be fitted against $(|\gamma_n|^2)_n$ relative to $\Phi_0, \Phi_1, \varphi$ and $\beta$. The resulting best-fit parameters would give a more physically correct law $\psi_n^\text{fit} = \text{arg}(\gamma_n^\text{fit})$ than what is achieved with usual algorithms before resorting to line-by-line corrections. This method could in the future prove faster and more accurate than the arbitrary polynomial phase-fitting protocol.

The derivation of $\hat{E}_\text{out}$ given in Eq.~(\ref{Eq:E^out}) also enables a novel spectral optimization method: it is possible to exactly invert the spectrum, rendering it flat, and then to multiply it by a second arbitrary filter. Hence, one may reshape the spectrum into a Gaussian curve. This Gaussian reshaping method (GRM) consists in the multiplication of $\hat{E}_\text{out}$ with the $H$ filter:
\begin{equation}
\label{Eq: gauss-reshape H}
    H(\omega) = \sum_n\gamma_n^{-1}G_\Delta(\omega)\delta_n(\omega)\ \text{with}\  G_\Delta(\omega) = e^{-\frac{(\omega - \omega_0)^2}{(\Delta\omega_m)^2}} 
\end{equation}

Modern pulse shapers reach dynamic attenuation ranges superior to 30~dB, however, this finite attenuation may still alter the desired complex SPD filter $F = |H^2|\cdot\text{arg}(H)$. In the following, special care is taken to emulate a realistic device: first, only the spectral modes with a power larger than 1\% of the maximum are reshaped, to ensure that the transmitted power remains sizable. The selected lines are reshaped with a dynamic range of 35~dB attenuation, while non-selected lines are put to 0. $F$ is then scaled to 1 in magnitude so that no artificial gain may be introduced. Finally, all lines of $|F|$ lower than 35~dB are supposed rerouted in the device and given an 50~dB attenuation.

Let us emphasize that the GRM, contrary to a simple Gaussian filtering method (GFM), completely cancels the ripples of the spectrum and gives a real-valued Gaussian curve truncated at $\pm\beta$ values. The range of $\Delta$ is experimentally limited by $\beta$: when $\Delta/\beta \ll 1$, a large portion of the EOML spectrum is filtered out, hence leading to much longer pulses. On the contrary for increasing $\Delta/\beta$, the resulting spectrum more and more resembles a square function of width $2\beta$. The main pulse is shortened but side lobes scaled as $\text{sinc}^2(\pi t / 2\beta T_m)$ lead to a degradation of the extinction ratio. The range $ 0.5\leqslant \Delta/\beta\leqslant 1.5$ seems typically advisable. For the lower values of this interval, one may of course expect a non-negligible increase in pulse duration (typ. a factor 2 for $\Delta/\beta = 0.5$).

In order to assess the interest of GRM versus GFM in terms of peak power, we normalize all time profiles by their energy, defined here as the integral value of the intensity over one period $T_m$. This ensures a proper comparison between the different cases without any dependence on the severity of the filter. Experimentally, this would for instance correspond to a setup where the EOML is launched after pulse shaping in a saturated amplifier. We hereby name this operation "energy-normalization".

\begin{figure}[!ht]
\centering
\includegraphics[width=\columnwidth]{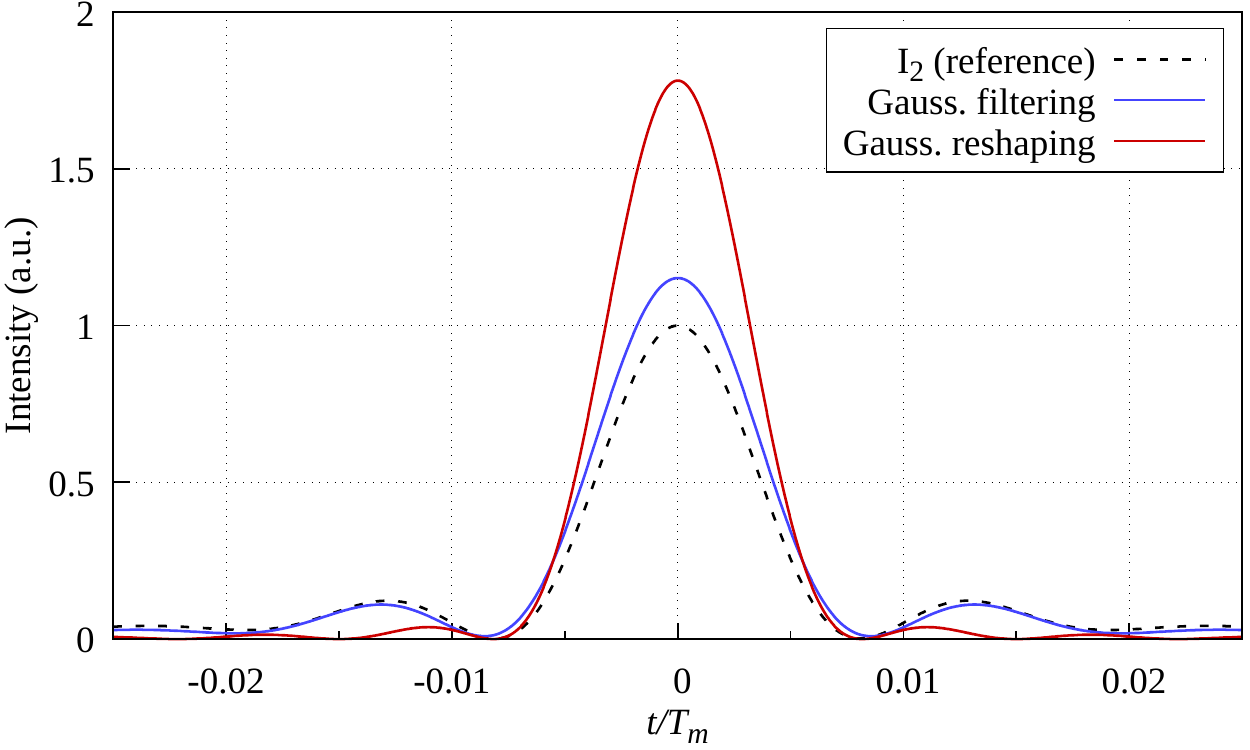}
\caption{\label{fig:Gauss_reshape_ppower} Intensity profiles of $I_2$ (reference, dashed), $I_2$ after GFM (blue), or GRM (red). Both filters are computed with $\Delta/\beta =1.5$. All profiles are energy-normalized and scaled so that the reference has a magnitude of 1.}
\end{figure}

The case $\Delta/\beta = 1.5$ is illustrated in Fig. \ref{fig:Gauss_reshape_ppower} and presents an interesting feature: the peak intensity after GRM is much higher and the pulse duration unchanged. GRM hence increases the energy in the main peak by a sizable amount of 75\%. In comparison, GFM only increases the pulse energy by 14\%.

\begin{figure}[!ht]
\centering
\includegraphics[width=\columnwidth]{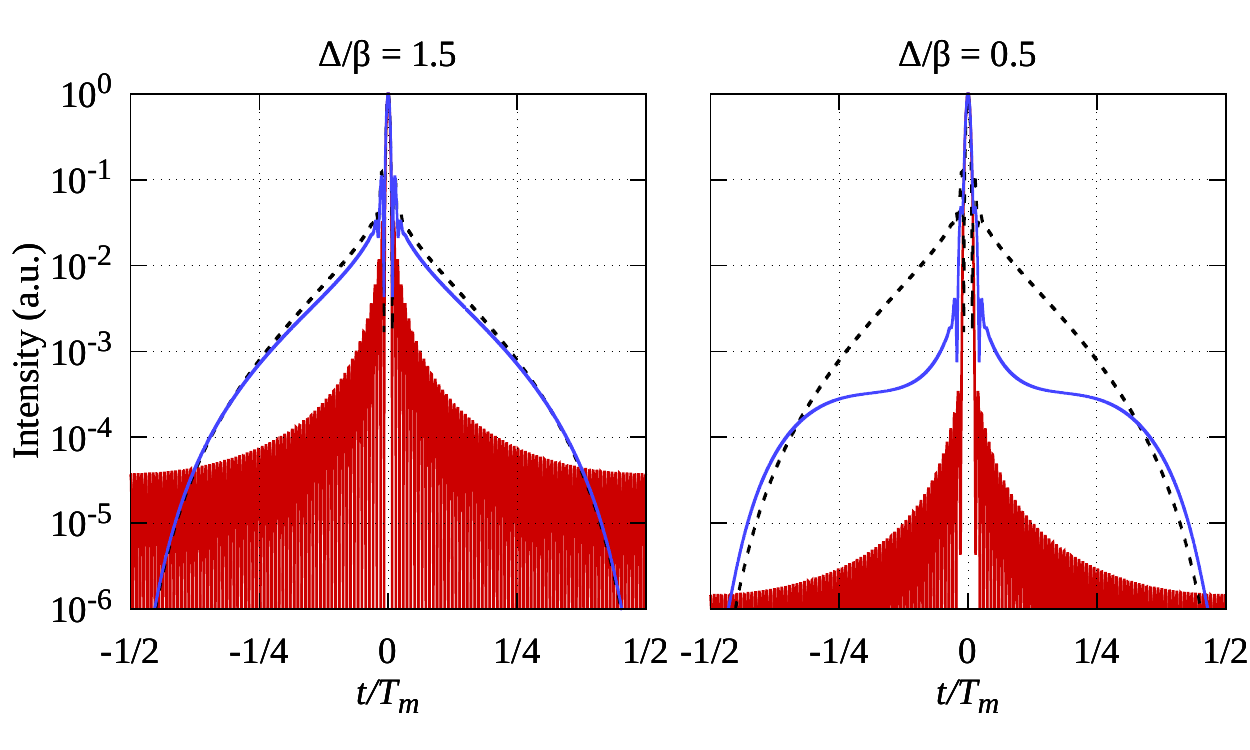}
\caption{\label{fig:Gauss_reshape_contrast}Comparison of $I_2$ (reference, dashed), GFM (blue) and GRM (red) profiles (all scaled to 1 in magnitude).} 
\end{figure}

The potential of GRM in terms of extinction ratio is  demonstrated in Fig. \ref{fig:Gauss_reshape_contrast}. In the worst case ($\Delta/\beta=1.5$), an enhancement over $I_\text{FL}$ of more than one order of magnitude is realized. For $\Delta/\beta=0.5$, the extinction ratio drops between $10^{-6}$ (half-period after the pulse) and $3\times10^{-3}$ (first satellite peak), values that are comparable to those of fiber mode-locked lasers \citep{Stuart:16}. 

As many studies have already proven it  \citep{chatellus2013,bolger:05,jannson:81,azana:04}, it is possible to multiply the repetition rate of a periodic optical signal with an integer factor by carrying out phase-only line-by-line shaping. This remarkable phenomenon is called the temporal Talbot effect. Starting from a periodic pulsed signal with a flat spectral phase, adding to each mode $n$ the quadratic phase term $\psi_T(n) = \pi n^2/q\ (q \in\mathbb{N})$, multiplies the repetition rate by $q$. 

If the rate-multiplication has already been achieved optically with a pulse shaper \cite{Caraquitena:07}, it has never, to the best of our knowledge, been tried on an EOML. Indeed, we have seen above (cf Fig.~\ref{fig:eoml_contrast}) how its textured spectrum provokes a slow time decrease of the intensity, precluding large multiplication factors. Temporal self-imaging is also limited by the bandwidth \cite{chatellus2013}: when $q \geqslant \beta$, the pedestals of the pulses overlap, leading to important pulse-to-pulse variations and a decrease of the extinction ratio. With a non-negligible residual phase, this defects happen for even smaller $q$ values, as small variations about $\psi_T$ induce sizable artefacts temporally \cite{Caraquitena:07}. 

\begin{figure}[!ht]
\centering
\includegraphics[width=\columnwidth]{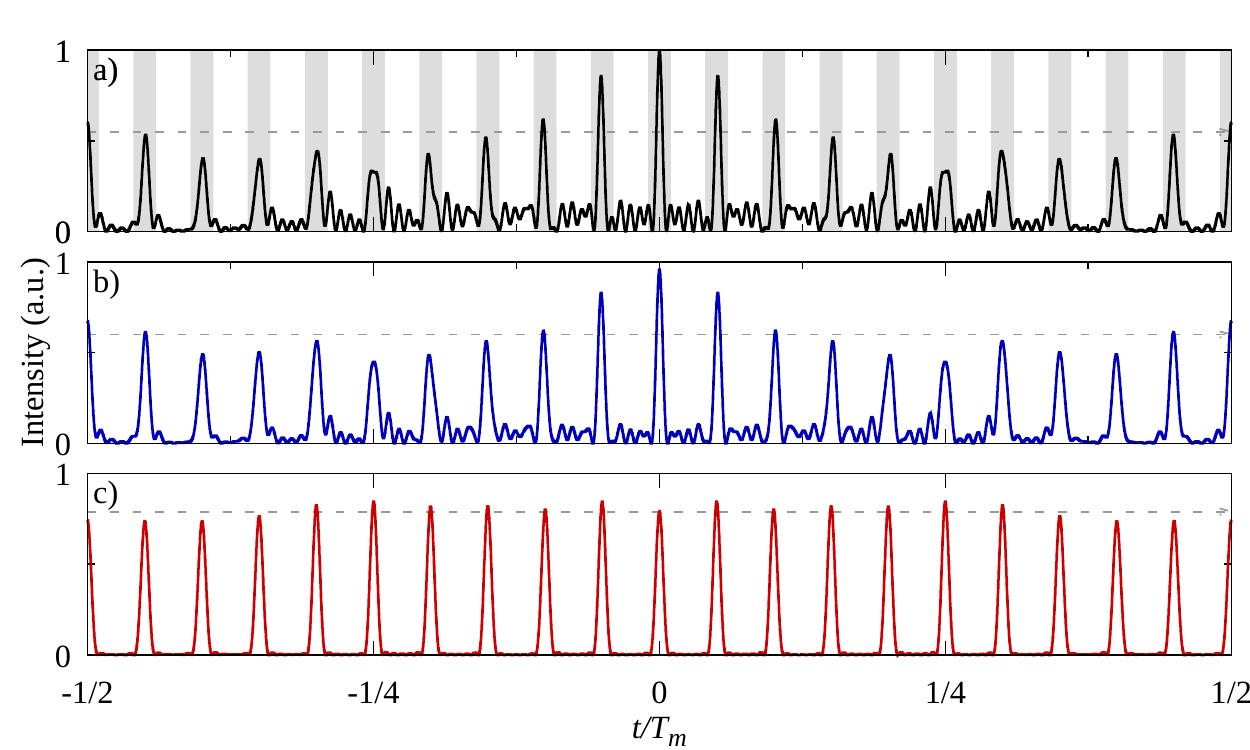}
\caption{\label{fig:talbot_comp}Talbot rate-multiplication ($q = 20$) obtained in three different cases: a) $\hat{E}_2$, b) GFM, c) GRM. b) and c) are computed with $\Delta/\beta = 0.85$; all profiles are energy-normalized, and scaled so that a) has a magnitude of 1.}
\end{figure}

This perturbation is illustrated for EOMLs in Fig. \ref{fig:talbot_comp}, which shows the intensity obtained when applying a Talbot phase $\psi_T$ with $q=20$ on different EOML spectra. The reference case (Fig.~\ref{fig:talbot_comp}a) presents the profile given by $E_2$, as would be observed with only a quadratic compression. For Fig.~\ref{fig:talbot_comp}b, a GFM with $\Delta/\beta = 0.85$ is applied on $\hat{E}_2$ before applying $\psi_T$. Finally (Fig.~\ref{fig:talbot_comp}c), one carries out the GRM of same width $\Delta$ as b).

In all cases, the rate multiplication is evident, with 20 peaks in one period $T_m$, but the reference signal shows a clear degradation. The GFM improves partly the peak-to-peak stability and the extinction ratio. The GRM cures these defects almost completely, with a peak-to-peak variation of 7\% and an extinction ratio about $5\times10^{-3}$. By summing the energy dispatched in the gray time-windows of Fig. \ref{fig:talbot_comp}, we can establish a total energy increase in the peaks superior to 40\% for the case of GRM, compared to the reference case a). This is the  consequence of the temporal redistribution of the energy by proper spectral reshaping.

To conclude, we have derived the general expression of the complex spectrum of EOMLs and characterized their time contrast, which is many orders of magnitude above that of mode-locked lasers. We then showed how pulse shapers could be put to interesting and optimal use: first, the exact knowledge of the spectrum may improve the phase-fitting procedure. Second, the Gaussian reshaping method may greatly improve both the peak power and the time contrast of EOMLs. Finally, the GRM coupled to Talbot effect enables large rate-multiplication factors.

As a prospective, we note that the presented results allow to envision very agile sources with about 100~fs pulse duration and a very wide range of repetition rate. Considering that modulators at 1500~nm may be driven at 80 GHz and, properly cascaded \cite{metcalf2019}, can generate spectra typically up to $\beta = 50\pi$, EOMLs implemented with the described method would have a repetition rate adjustable between 80~GHz and 5~THz ($q = 63 <\beta/2$). This new class of laser sources should have interesting applications in diverse fields such as telemetry, spectroscopy or THz generation.

\section*{Acknowledgments}
The authors wish to thank Prof. J.-C.~Delagnes and Dr.~G.~Duchateau for insightful discussions.

\section*{Disclosures}
The authors declare no conflicts of interest. This research did not receive any specific grant from funding agencies in the public, commercial, or not-for-profit sectors.

\bibliography{eoml}

\end{document}